\documentclass[%showpacs,
showkeys,12pt,
preprint,preprintnumbers,nofootinbib,
groupedaddress,superscriptaddress,amsmath,amssymb]{revtex4}
\usepackage{graphicx}% Include figure files
\usepackage{dcolumn}% Align table columns on decimal point
\usepackage{bm}% bold math
\usepackage{amssymb}
\usepackage{amsmath}
\usepackage{epsfig}    
\usepackage{color}
\usepackage{slashed}
\usepackage{hhline}
\def\be{\begin{equation}}
\def\ee{\end{equation}}
\newcommand{\bea}{\begin{eqnarray}}
\newcommand{\eea}{\end{eqnarray}}
\newcommand{\nn}{\nonumber}

\numberwithin{equation}{section}

\begin{document}

{\begin{flushright}{KIAS-P16046}
\end{flushright}}

%%%%%%%%%
\title{ Radiatively induced Quark and Lepton Mass Model}
%\preprint{KIAS-P14078}
%

\author{Takaaki Nomura}
\email{nomura@kias.re.kr}
\affiliation{School of Physics, KIAS, Seoul 130-722, Korea}

\author{Hiroshi Okada}
\email{macokada3hiroshi@gmail.com}
\affiliation{Physics Division, National Center for Theoretical Sciences, Hsinchu, Taiwan 300}

\date{\today}

\begin{abstract}
We  propose a radiatively induced quark and lepton mass model in the first and second generation with extra $U(1)$ gauge symmetry and vector-like fermions. Then we analyze the allowed regions which simultaneously satisfy the FCNCs for the quark sector, LFVs including $\mu-e$ conversion, the quark mass and mixing, and the lepton mass and mixing. Also we estimate the typical value for the $(g-2)_\mu$ in our model.
\end{abstract}
\maketitle
\newpage

\section{Introduction}
Radiatively induced mass models are one of the promising candidate to include a dark matter (DM) candidate naturally, which connect the standard model (SM) fermions and DM. Along this line of idea, there exist a lot of papers, {\it i.e.}, ~\cite{onelps} at one-loop level, ~\cite{twolps} at two-loop level,~\cite{threelps} at three-loop level, and~\cite{Fourlps} at four-loop level.
%%%%%%%
However authors mainly focus on the neutrino sector, but not so many on the quark sector~\cite{Balakrishna:1987qd, Balakrishna:1988bn, Ma:2013mga,Ma:2014yka, Kownacki:2016hpm,Arbelaez:2016mhg,Ma:1989tz,He:1989er, Ibarra:2014fla, Baumgart:2014jya}. 

In this paper, we propose all the SM fermion masses are induced at one-loop level except for the third generation, and the inert type of boson DM couples to all these fermions. 
{The masses of third generation fermions are generated via the vacuum expectation value (VEV) of SM Higgs field to be consistent with SM Higgs properties observed by LHC experiments such as 
gluon fusion cross section and $h \to b \bar b (\tau \bar \tau)$ branching fractions~\cite{Aad:2015gba, CMS:2014ega}.
Furthermore it would be natural to require first and second generation masses are generated at loop-level from the fermion mass hierarchy.
Then we add extra local U(1) symmetry to restrict the Yukawa interaction associated with SM Higgs field in anomaly free way.
The vector-like fermions are also introduced to write relevant one-loop diagrams for fermion mass generation.
In our model, therefore, the light fermion masses are generated at the one-loop level induced by the Yukawa interactions among SM fermions, inert scalar fields and vector-like fermions which are invariant under the new U(1).
We note that all these Yukawa couplings cannot be so large to induce the relevant relic abundance of DM in our parameter choices, and the nature of DM is the same as the two Higgs doublet model with one inert $SU(2)_L$ doublet boson.}
In order to reproduce the observed mixing matrices and masses for the lepton and quark sector, we have to take into account the flavor changing neutral currents (FCNCs) and lepton flavor violations (LFVs) where 
 mediated boson masses (including DM) are restricted by the both sector. Also positive contribution to the anomalous magnetic moment 
are induced from the lepton sector via one-loop diagram in which the mediated boson are included. Therefore, an economical scenario including the quark sector might be achieved in a sense.  

This paper is organized as follows.
In Sec.~II, we show our model, %to introduce exotic fermions and bosons with some additional symmetries,  
and establish  the quark and lepton sector, and derive the analytical forms of  FCNCs, LFVs, muon anomalous magnetic moment.
In Sec.~III, we have a numerical analysis, and show some results.
We conclude and discuss in Sec.~IV.
%\newpage

%%%%%%%%%%%%%%%%%%%%%%%%%%%%%%%%%%%%%
%\section{The Model}
%\subsection{Model setup}

 \begin{widetext}
\begin{center} 
\begin{table}%[tbc]
%\begin{tiny}
\begin{tabular}{|c||c|c|c|c|c|c||c|c|c|c|c|c|}\hline\hline  
&\multicolumn{6}{c||}{Quarks} & \multicolumn{5}{c|}{Leptons} \\\hline
Fermions& ~$Q_L^\alpha$~ & ~$u_R^i$~ & ~$d_R^i$ ~ & ~$t_R$~ & ~$b_R$ ~ & ~$Q'^i_{L(R)}$ 
& ~$L_L^\alpha$~ & ~$e_R^i$~ & ~$\nu_R^i$~& ~$\tau_R$ ~ & ~$L'^i_{L(R)}$~ 
\\\hline 
$SU(3)_C$ & $\bm{3}$  & $\bm{3}$  & $\bm{3}$ & $\bm{3}$ &
 $\bm{3}$  & $\bm{3}$  & $\bm{1}$  & $\bm{1}$   & $\bm{1}$  & $\bm{1} $  & $\bm{1}$ \\\hline 
 %%%
 $SU(2)_L$ & $\bm{2}$  & $\bm{1}$  & $\bm{1}$ & $\bm{1}$ &
 $\bm{1}$  & $\bm{2}$  & $\bm{2}$  & $\bm{1}$   & $\bm{1}$  & $\bm{1}$   & $\bm{2}$ \\\hline 
 %%%
$U(1)_Y$ & $\frac16$ & $\frac23$  & $-\frac{1}{3}$ & $\frac23$  & $-\frac{1}{3}$ & $\frac{1}{6}$
 & $-\frac12$ & $-1$  & $0$ &  $-1$  &  $-\frac{1}{2}$ \\\hline
 %%%
 $U(1)_{R}$ & $0$ & $x$  & $-x$ & $0$  & $0$ & $0$  & $0$ & $-x$ & $x$  & $0$   & $0$ \\\hline
 %%%
$Z_2$ & $+$ & $+$  & $+$ & $+$
& $+$ & $-$  & $+$ & $+$ & $+$ & $+$ & $-$ \\\hline
%%%
%$\mathbb{Z}_2$ & $+$   & $-$  & $+$ & $+$& $-$& $-$& $+$ & $+$  \\\hline\hline
\end{tabular}
\caption{Field contents of fermions
and their charge assignments under $SU(2)_L\times U(1)_Y\times U(1)_R\times Z_2$, where each of the flavor index is defined as $\alpha\equiv 1-3$ and $i=1,2$.}
\label{tab:1}
% \end{tiny}
\end{table}
\end{center}
\end{widetext}

\begin{table}[thbp]
\centering {\fontsize{10}{12}
\begin{tabular}{|c||c|c||c|c|c|}\hline\hline
&\multicolumn{2}{c||}{VEV$\neq 0$} & \multicolumn{2}{c|}{Inert } \\\hline
  Bosons  &~ $\Phi$   ~ &~ $\varphi$     ~ &~ $\eta$   ~ &~ $S$ ~ \\\hline
$SU(2)_L$ & $\bm{2}$  & $\bm{1}$   & $\bm{2}$ &
 $\bm{1}$ \\\hline 
$U(1)_Y$ & $\frac12$  & $0$ & $\frac{1}{2}$ & $0$   \\\hline
 $U(1)_R$ & $0$ & $x$ & $x$ & $0$   \\\hline
$Z_2$ & $+$ & $+$ & $-$ & $-$ \\\hline
\end{tabular}%
} 
\caption{Boson sector }
\label{tab:2}
\end{table}

\section{ Model setup}
In this section, we show our model. 
As for the fermion sector, we introduce $SU(2)_L$ doublet exotic vector-like fields $Q'\equiv [u',d']^T$ and $L'\equiv[N',E']^T$ with two flavors, where each of $Q'$  and $L'$ has triplet and singlet under $SU(3)_C$ and $Z_2$ symmetry are imposed. Also we introduce two right-handed neutrinos $\nu_R^i$ $(i=1,2)$, which constitute Dirac fields after the spontaneous electroweak symmetry breaking that are the same as the other three sectors in SM.
Then we impose an additional gauged $U(1)_R$ symmetry, where only the first and second family with right-handed SM fermions and $\nu_R^i$ have none-zero charge $x$.
Field contents and their assignments are summarized in Table~\ref{tab:1}, in which $i=1,2$ and  $\alpha=1-3$ represent the number of  family, and no index fields represent the third family.
%%%
{Notice here that we require the third generation couple to the SM-like Higgs directly for consistency with SM Higgs properties observed by the LHC experiments such as gluon fusion production cross section and branching fractions.}
%%%

As for the boson sector, we add two $SU(2)_L$ singlets $\varphi$ and $S$, and one $SU(2)_L$ doublet scalar $\eta$ to the Higgs-like boson $\Phi$,
 where $\Phi$ and $\varphi$ have the VEVs, symbolized by $\langle\Phi\rangle\equiv v/\sqrt2$ and $\langle\varphi\rangle\equiv v'/\sqrt2$,
that spontaneously break the electroweak and $U(1)_R$ symmetry respectively.  On the other hand, $S$ and $\eta$ do not have VEVs that are assured by the $Z_2$ symmetry.
Field contents and their assignments are summarized in Table~\ref{tab:2}, where we assume $S$ to be a real field for simplicity.

%%%%%%%%% Anomaly cancellation %%%%%%%%%
%Before proceeding our concrete analysis, we show how the anomaly cancellation occurs.
{\it Anomaly cancellation}:
%Here we have to check the anomaly cancellation among fermions, since $U(1)_R$ is gauged symmetry. 
The $U(1)_R$ gauge symmetry is anomaly free where the anomaly is canceled within each generation of fermions~\cite{Ko:2012hd}.
%Our anomaly actually vanishes each of generation, where the third generation is the same as the SM one.
We then assign $U(1)_R$ charges to first and second generation fermions but charges for third generation fermions are required to be zero.
The triangle anomaly within one generation cancels as follows:
\begin{align}
& [U(1)_Y]^2U(1)_R:\quad 3\left(\frac49x+\frac19(-x)\right)-x=0,\\
& [U(1)_R]^2U(1)_Y:\quad 3\left(\frac23x^2-\frac13 x^2\right)-x^2=0,\\
& [U(1)_R]^3:\quad 3\left(x^3-x^3\right)-x^3+x^3=0,\\
& U(1)_R:\quad 3\left(x-x\right)-x+x=0.
\end{align}
%%%%%%%%% %%%%%%%%% %%%%%%%%%

{\it Yukawa Lagrangian}:
Under these fields and symmetries, the renormalizable Lagrangians for quark and lepton sector are given by 
\begin{align}
-{\cal L}_{Q}&=(y_u)_{i j}\bar Q'_i u_{R_j} (i\sigma_2)\eta^* + (y_d)_{ij} \bar Q'_{ij}\eta d_{R_j} + (y_Q)_{\alpha j}\bar Q_\alpha Q'_j S\nn\\
&+(y_t)_{\alpha}\bar Q_\alpha t_{R_j} (i\sigma_2)\Phi^* + (y_b)_{\alpha} \bar Q_{\alpha}\Phi b_{R}
+m_{Q'_k} \bar Q'_k Q'_k
+{\rm c.c.}, \label{eq:lag-quark}\\
-{\cal L}_{L}&=(y_\nu)_{i j}\bar L'_i \nu_{R_j} (i\sigma_2)\eta^* + (y_\ell)_{ij} \bar L'_{ij}\eta e_{R_j} + (y_L)_{\alpha j}\bar L_\alpha L'_j S\nn\\
&+(y_\tau)_{\alpha}\bar L_\alpha \tau_{R_j} (i\sigma_2)\Phi + (y_{\nu_\tau})_{\alpha} \bar L_{\alpha}\Phi \nu_{\tau R}
+m_{L'_k} \bar L'_k L'_k+{\rm c.c.},
\label{eq:lag-quark}
\end{align}
where $\sigma_2$ is the Pauli matrix.

%%%%%%%%%%%%%%%%%%%
{\it Higgs potential}:
Higgs potential is given by
\begin{align}
V&=
m_\varphi^2 |\varphi|^2 + m^2_S S^2+ m_\Phi^2 |\Phi|^2+m_\eta^2 |\eta|^2  +\lambda_0 \left[(\Phi^\dag\eta)S\varphi^* +{\rm c.c.} \right]\\
%%%%
&
+\lambda_\varphi|\varphi|^4+\lambda_S S^4 + \lambda_\Phi |\Phi|^4+\lambda_\eta |\eta|^4
+\lambda_{\varphi S} |\varphi|^2 S^2 +\lambda_{\varphi \Phi} |\varphi|^2 |\Phi|^2+\lambda_{\varphi \eta} |\varphi|^2|\eta|^2\nn\\
&
+\lambda_{S \Phi} S^2 |\Phi|^2 + \lambda_{S \eta} S^2|\eta|^2 
+\lambda_{\Phi\eta}  |\Phi|^2 |\eta|^2+\lambda'_{\Phi\eta}|\Phi^\dag\eta|^2, \label{eq:lag-pot}
\end{align}
where the scalar fields are parameterized as 
\begin{align}
%\begin{tiny}
&\Phi =\left[
\begin{array}{c}
w^+\\
\frac{v+\phi+iz}{\sqrt2}
\end{array}\right],\quad 
%%%
\eta =\left[
\begin{array}{c}
\eta^+\\
\frac{\eta_R+i\eta_I}{\sqrt2}
\end{array}\right],
\quad
\varphi=\frac{v'+\varphi_R+i\varphi_I}{\sqrt2},
\label{component}
%\end{tiny}
\end{align}
where $w^\pm$, $z$, and $\varphi_I$ are respectively absorbed by the longitudinal degrees of freedom of charged SM gauged boson $W^\pm$, neutral SM gauged $Z$, and neutral $U(1)_R$  gauged boson $Z'$.
After the spontaneous symmetry breaking, neutral bosons mix each other as follows:
\begin{align}
\left[
\begin{array}{c}
S\\ \eta_R
\end{array}\right]
={\cal O}_\alpha
\left[
\begin{array}{c}
H_1\\ H_2
\end{array}\right],\quad
%%%%%%%
\left[
\begin{array}{c}
\varphi_R\\ \phi
\end{array}\right]
={\cal O}_ \beta
\left[
\begin{array}{c}
h_1\\ h_2
\end{array}\right],\quad
{\cal O}_ a
\equiv
\left[
\begin{array}{cc}
c_a & s_a\\ 
-s_a & c_a \\
\end{array}\right],
\end{align}
where we define $c_a\equiv \cos a$, $s_a\equiv \sin a$,  $H_i(i=1,2)$ is the mass eigenstate of the inert neutral boson, and
 $h_i(i=1,2)$ is the mass eigenstate of the neutral boson with VEVs. Here $h_2$ is the SM-like Higgs and $h_1$ is the additional Higgs boson (like a 750 GeV boson). All of the mass eigenvalues and mixings are written in terms of VEVs, and quartic couplings in the Higgs potential after inserting the tadpole conditions: $\partial V/\partial \phi|_{v,v'}=0$ and $\partial V/\partial \varphi_R|_{v,v'}=0$. 
% \textcolor{red}{If you like, please write each of values explicitly...}
 
 %%%%%%%%%%%%%%%%%%%
 {
{\it Z' boson}:
After $U(1)_R$ symmetry breaking by VEV of $\varphi$, we have massive $Z'$ boson where the mass is $m_{Z'} = x g_R v'$ ($g_R$ is gauge coupling for $U(1)_R$).
The $Z'$ couples to right-handed SM fermions at tree level since first and second generation right-handed fermions have $U(1)_R$ charge:
\begin{align}
{\cal L} \supset & g_R x Z'_\mu \sum_{k=1,2} \Bigl( \bar  u^\alpha_R (V_{u_R})^\dagger_{\alpha k} (V_{u_R})_{k \beta} \gamma^\mu u_R^\beta - \bar d^\alpha_R (V_{d_R})^\dagger_{\alpha k} (V_{d_R})_{k\beta} \gamma^\mu d_R^\beta  \nonumber \\
& \qquad \qquad - \bar e^\alpha_R (V_{e_R})^\dagger_{\alpha k} (V_{e_R})_{k\beta} \gamma^\mu e_R^\beta + \bar \nu^\alpha_R (V_{\nu_R})^\dagger_{\alpha k} (V_{\nu_R})_{k\beta} \gamma^\mu \nu_R^\beta  \Bigr),
\end{align}
where $\{\alpha,\beta \} = 1,2,3$ and $V_{f_R}$ are unitary matrices for diagonalizing fermion mass matrices. Note that the matrices $\sum_k (V_{f_R})^\dagger_{\alpha k} (V_{f_R})_{k \beta}$ are not unity in general since only first and second fermions have $U(1)_R$ charge. 
Thus we have flavor changing interaction in $Z'$ exchange.
Since $Z'$ couples to both quarks and leptons the mass is strictly constrained by dilepton search at the LHC; $m_{Z'} \gtrsim 3$ TeV~\cite{Chatrchyan:2012oaa, Aad:2014cka, ATLAS:dilepton} if order of $g_R$ is the same as SM gauge coupling.
In this paper, we do not further discuss the $Z'$ since it is not relevant for light fermion mass generation and 
mass of $Z'$ is assumed to be sufficiently heavy so that it does not affect flavor constraints. }
 
 \subsection{Quark sector}
In this subsection, we will analyze the quark sector.
First of all, let us focus on the Yukawa sector, in which the measured SM quark masses and their mixings are induced.~\footnote{An interesting idea to generate the quark masses and mixings has been discussed in Ref.~\cite{Altmannshofer:2014qha} in the framework of supersymmetry.
Here these mass spectrum and their mixings are induced through the renormalization equations, starting from only the third generation.
See also Ref.~\cite{Ibarra:2014pfa} for the lepton sector.}
Up and down quark mass matrices are diagonalized by $M^{diag.}_u=V_{u_L} M_u V_{u_R}$, and $M^{diag.}_d=V_{d_L} M_d V_{d_R}$,
where $V's$ are unitary matrix to give their  diagonalization matrices. Then CKM matrix is defined by $V_{CKM}\equiv V^\dag_{u_L }V_{d_L}$, where
it can be parametrized by three mixings with one phase as follows:
\begin{align}
V_{CKM}\equiv V^\dag_{u_L }V_{d_L} \equiv
%%%
\left[
\begin{array}{ccc}
1 & 0 & 0 \\ 
0 & c_{23} & s_{23} \\
0 & -s_{23} & c_{23} \\
\end{array}\right]
%%%
\left[
\begin{array}{ccc}
c_{13} & 0 & s_{13}e^{-i\delta} \\ 
0 & 1 & 0 \\
-s_{13}e^{i\delta}  & 0 & c_{13} \\
\end{array}\right]
\left[
\begin{array}{ccc}
c_{12} & s_{12}e^{-i\delta} & 0 \\ 
-s_{12} & c_{12} & 0 \\
0  & 0 & 1 \\
\end{array}\right]
,
\end{align}
 where in the numerical analysis, we assume to take the following forms to evade the stringent constraint of $B_0-\bar B_0$ mixing in the numerical analysis:
  \begin{align}
 V^\dag_{u_L }= \left[
\begin{array}{ccc}
1 & 0 & 0 \\ 
0 & c_{23} & s_{23} \\
0 & -s_{23} & c_{23} \\
\end{array}\right]
%%%
\left[
\begin{array}{ccc}
c_{13} & 0 & s_{13}e^{-i\delta} \\ 
0 & 1 & 0 \\
-s_{13}e^{i\delta}  & 0 & c_{13} \\
\end{array}\right],\quad
%%%
V_{d_L }= \left[
\begin{array}{ccc}
c_{12} & s_{12}e^{-i\delta} & 0 \\ 
-s_{12} & c_{12} & 0 \\
0  & 0 & 1 \\
\end{array}\right].
 \end{align}
 The mass matrix in our form is written in terms of tree level mass matrix  and one-loop one as
 \begin{align}
(M_{u(d)})_{\alpha\beta}=(M^{tree}_{u(d)})_{\alpha 3} + (M^{one-loop}_{u(d)})_{\alpha j},
 \end{align}
 with
  \begin{align}
&   (M^{tree}_{u(d)})_{\alpha 3} = \frac{(y_{t(b)})_\alpha}{\sqrt2},\\
%(m_d)_{\alpha\beta}=(m^{tree}_d)_{\alpha j}+ (m^{one-loop}_d)_{\alpha 3},\\
&(M^{one-loop}_{u(d)})_{\alpha j}=s_\alpha c_\alpha
\sum_{k=1,2}\frac{(y_Q)_{\alpha k} m_{Q'_k}(y_{u(d)})_{kj} }{\sqrt2(4\pi)^2}
\int_0^1dx_1
\frac{x_1 m^2_{Q'_k}+(1-x_1)m^2_{H_1} }{x_1 m^2_{Q'_k}+(1-x_1)m^2_{H_2} }.
 \end{align}
 
 {\it Flavor changing neutral currents}:
 Now we discuss the constraints on the quark sector. The stringent constraints come from the flavor changing neutral currents (FCNCs), which are called $Q-\bar Q$ mixing and given in terms of the mass difference between a meson and an anti-meson. Here we symbolize these observables as $\Delta m_Q$ with $Q=D,K,B$. Then each of our  formulae is given at box-type one-loop level by~\cite{Gabbiani:1996hi} 
 \begin{align}
 & \Delta m_D \approx %_{III}
 %-\frac{5}{24}\left(\frac{m_D}{m_c+m_u}\right)^2m_Df_D^2 
\frac{1}{2(4\pi)^2} \sum_{\rho,\sigma}^{1,2} 
 \left(
{\cal F}_{\rho\sigma}^- C^D_{Q_2} ({\cal M}_{Q_2}^D)_{\rho\sigma}^2% (y'_u)_{\rho2}m_{Q'_\rho}(y'_{Q})_{1\rho} (y'_{Q})_{1\sigma}m_{Q'_\sigma}(y'_u)_{\sigma2}
 +
{\cal F}_{\rho\sigma}^+ C^D_{Q_4} ({\cal M}_{Q_4}^D)_{\rho\sigma}^2
 % (y'_u)_{\rho2}m_{Q'_\rho}(y'_{Q})_{1\rho} (y'_{u})^\dag_{1\sigma} m_{Q'_\sigma} (y'_{Q})^\dag_{\sigma2} 
 \right),
% \nn\\ & \left[ \frac{s_{2R}^2}2 (F_1[H_1] + F_1[H_2]) + c_{2R}^2  (F_2[H_1,H_2] + F_1[H_2,H_1])-c_R^2 F_2[\eta_I,H_1] -s_R^2 F_2[\eta_I,H_2] \right]
%%%% %%%
\\
 & \Delta m_K \approx  %_{III}
 %-\frac{5}{24}\left(\frac{m_D}{m_c+m_u}\right)^2m_Df_D^2 
\frac{1}{2(4\pi)^2} \sum_{\rho,\sigma}^{1,2}
 \left( 
{\cal F}_{\rho\sigma}^- C^K_{Q_2} ({\cal M}_{Q_2}^K)_{\rho\sigma}^2% (y'_u)_{\rho2}m_{Q'_\rho}(y'_{Q})_{1\rho} (y'_{Q})_{1\sigma}m_{Q'_\sigma}(y'_u)_{\sigma2}
 +
{\cal F}_{\rho\sigma}^+ C^K_{Q_4} ({\cal M}_{Q_4}^K)_{\rho\sigma}^2
 % (y'_u)_{\rho2}m_{Q'_\rho}(y'_{Q})_{1\rho} (y'_{u})^\dag_{1\sigma} m_{Q'_\sigma} (y'_{Q})^\dag_{\sigma2} 
 \right) ,
 %%%% %%%
\\
 & \Delta m_B \approx  %_{III}
 %-\frac{5}{24}\left(\frac{m_D}{m_c+m_u}\right)^2m_Df_D^2 
\frac{1}{2(4\pi)^2} \sum_{\rho,\sigma}^{1,2}
 \left(  
{\cal F}_{\rho\sigma}^- C^B_{Q_2} ({\cal M}_{Q_2}^B)_{\rho\sigma}^2% (y'_u)_{\rho2}m_{Q'_\rho}(y'_{Q})_{1\rho} (y'_{Q})_{1\sigma}m_{Q'_\sigma}(y'_u)_{\sigma2}
 +
{\cal F}_{\rho\sigma}^+ C^B_{Q_4} ({\cal M}_{Q_4}^B)_{\rho\sigma}^2
 % (y'_u)_{\rho2}m_{Q'_\rho}(y'_{Q})_{1\rho} (y'_{u})^\dag_{1\sigma} m_{Q'_\sigma} (y'_{Q})^\dag_{\sigma2} 
 \right) ,
 \end{align}
 with
  \begin{align}
&({\cal M}_{Q_2}^D)_{\rho\sigma}^2 \equiv (y'_u)_{\rho2}m_{Q'_\rho}(y'_{Q})_{1\rho} (y'_{Q})_{1\sigma}m_{Q'_\sigma}(y'_u)_{\sigma2},\
({\cal M}_{Q_4}^D)_{\rho\sigma}^2 \equiv  (y'_u)_{\rho2}m_{Q'_\rho}(y'_{Q})_{1\rho} (y'_{u})^\dag_{1\sigma} m_{Q'_\sigma} (y'_{Q})^\dag_{\sigma2},\\
&({\cal M}_{Q_2}^K)_{\rho\sigma}^2 \equiv (y'_d)_{\rho2}m_{Q'_\rho}(y'_{Q})_{1\rho} (y'_{Q})_{1\sigma}m_{Q'_\sigma}(y'_d)_{\sigma2},\
({\cal M}_{Q_4}^K)_{\rho\sigma}^2 \equiv  (y'_d)_{\rho2}m_{Q'_\rho}(y'_{Q})_{1\rho} (y'_{d})^\dag_{1\sigma} m_{Q'_\sigma} (y'_{Q})^\dag_{\sigma2},\\
&({\cal M}_{Q_2}^B)_{\rho\sigma}^2 \equiv (y'_d)_{\rho3}m_{Q'_\rho}(y'_{Q})_{1\rho} (y'_{Q})_{1\sigma}m_{Q'_\sigma}(y'_d)_{\sigma3},\
({\cal M}_{Q_4}^B)_{\rho\sigma}^2 \equiv  (y'_d)_{\rho3}m_{Q'_\rho}(y'_{Q})_{1\rho} (y'_{d})^\dag_{1\sigma} m_{Q'_\sigma} (y'_{Q})^\dag_{\sigma3},\\
&C^D_{Q_2}=-\frac{5}{24}\left(\frac{m_D}{m_c+m_u}\right)^2m_D f_D^2,\quad
C^D_{Q_4}=\left[\frac{1}{24}+\frac{1}{4} \left(\frac{m_D}{m_c+m_u}\right)^2\right]m_D f_D^2,\\
%%%
&C^K_{Q_2}=-\frac{5}{24}\left(\frac{m_K}{m_s+m_d}\right)^2m_K f_K^2,\quad
C^K_{Q_4}=\left[\frac{1}{24}+\frac{1}{4} \left(\frac{m_K}{m_s+m_d}\right)^2\right]m_K f_K^2,\\
%%%
&C^B_{Q_2}=-\frac{5}{24}\left(\frac{m_B}{m_b+m_d}\right)^2m_B f_B^2,\quad
C^B_{Q_4}=\left[\frac{1}{24}+\frac{1}{4} \left(\frac{m_B}{m_b+m_d}\right)^2\right]m_B f_B^2,
  \end{align}
and 
\begin{align}
&{\cal F}_{\rho\sigma}^\pm \equiv \left[ \frac{s_{2R}^2}2 (F_1[H_1] + F_1[H_2]) + c_{2R}^2  (F_2[H_1,H_2] + F_1[H_2,H_1]) \pm c_R^2 F_2[\eta_I,H_1] \pm s_R^2 F_2[\eta_I,H_2] \right]_{\rho\sigma},\\
%%%
&F_1[m_a]\equiv \int_0^1da  \int_0^{1-a}db % \int_0^{1-a-b} dc
\frac{1-a-c}{\left[a m_{Q'_\rho}^2 + b m_{Q'_\sigma}^2+(1-a-b) m_a^2 \right]^2},\\
%%%
&F_2[m_a,m_b]\equiv \int_0^1da  \int_0^{1-a}db  \int_0^{1-a-b} dc
\frac{1}{\left[a m_{Q'_\rho}^2 + b m_{Q'_\sigma}^2+c m_a^2+(1-a-b-c) m_b^2 \right]^2},
 \end{align}
 where the Yukawa couplings are defined to be $y'_{u(d)} \equiv y_{u(d)} V^\dag_{u(d)_R}$, $y'_Q\equiv V_{u_L} y_Q$, and assumed to be $V_L=V_R$ in our analytical convenience. 
%%%
Experimental and input values~\cite{Okada:2016whh} are given  by
\begin{align} 
&m_u\approx 2.3 [{\rm MeV}],\ m_c\approx 1275 [{\rm MeV}],\ m_t\approx 173.2 [{\rm GeV}],\\ 
&m_d\approx 4.8 [{\rm MeV}],\ m_s\approx 95 [{\rm MeV}],\ m_t\approx 4.18 [{\rm GeV}],\\ 
%%%
& m_D\approx 1864.84 [{\rm MeV}],\ m_K \approx 497.614 [{\rm MeV}],\ m_B \approx 5279.50 [{\rm MeV}],\\
& f_D\approx 212  [{\rm MeV}],\ f_K\approx 159.8  [{\rm MeV}],\ f_B\approx 200  [{\rm MeV}].
 \end{align}
Finally the experimental upper bounds are respectively given by~\cite{Okada:2016whh}
 \begin{align} 
&\Delta m_D\lesssim 6.25\times 10^{-12}[{\rm MeV}],\\ 
&\Delta m_K\lesssim 3.484\times 10^{-12}[{\rm MeV}],\\ 
&\Delta m_B\lesssim 3.356\times 10^{-10}[{\rm MeV}].
 \end{align}

  \subsection{Lepton sector}
In this subsection, we will discuss the lepton sector, where neutrinos are supposed to be Dirac neutrino.
Thus the process to induce the mass matrix in the lepton sector is the same as the quark sector except the third generation of the neutrino.
So we just provide  the definitions by changing 
$u\to\nu$, $d\to \ell$, $V_{CKM}\to V_{MNS}$  in the quark sector . 
 %%%
 Then the leptons mass matrix in our form is written as
 \begin{align}
(M_{\nu(\ell)})_{\alpha\beta}=(M^{tree}_{\ell})_{\alpha 3} + (M^{one-loop}_{\nu(\ell)})_{\alpha j},
 \end{align}
 with
  \begin{align}
&   (M^{tree}_{\ell})_{\alpha 3} = \frac{(y_{\tau})_\alpha}{\sqrt2},\\
%(m_d)_{\alpha\beta}=(m^{tree}_d)_{\alpha j}+ (m^{one-loop}_d)_{\alpha 3},\\
&(M^{one-loop}_{\nu(\ell)})_{\alpha j}=s_\alpha c_\alpha
\sum_{k=1,2}\frac{(y_{\nu (L)})_{\alpha k} m_{L'_k}(y_{S})_{kj} }{\sqrt2(4\pi)^2}
\int_0^1dx_2
\frac{x_2 m^2_{L'_k}+(1-x_2)m^2_{H_1} }{x_2 m^2_{L'_k}+(1-x_2)m^2_{H_2} }.
 \end{align}

\begin{table}[t]
\begin{tabular}{c|c|c|c} \hline
Process & $(j,\alpha)$ & Experimental bounds ($90\%$ CL) & References \\ \hline
%%%%%%%
$\mu^{-} \to e^{-} \gamma$ & $(2,1)$ &
	${BR}(\mu \to e\gamma) < 4.2 \times 10^{-13}$ & \cite{TheMEG:2016wtm} \\
$\tau^{-} \to e^{-} \gamma$ & $(3,1)$ &
	${Br}(\tau \to e\gamma) < 3.3 \times 10^{-8}$ & \cite{Adam:2013mnn} \\
$\tau^{-} \to \mu^{-} \gamma$ & $(3,2)$ &
	${BR}(\tau \to \mu\gamma) < 4.4 \times 10^{-8}$ & \cite{Adam:2013mnn}   \\ \hline
$\mu \to e\ {\rm conversion}$ & $(2,1)$ &
	${R}(Ti) < 4.3 \times 10^{-12} \to{\cal O}(10^{-18})({\rm future\ bound })$ & \cite{Dohmen:1993mp}$\to$ \cite{Hungerford:2009zz, Cui:2009zz}   \\ \hline
\end{tabular}
\caption{Summary of $\ell_j \to \ell_\alpha \gamma$ process and the lower bound of experimental data.}
\label{tab:Cif}
\end{table}

 {\it Lepton flavor violations}:
 The lepton flavor (LFVs) violation processes give the constraints on our parameters.
 The most known processes are $\ell_j\to \ell_\alpha\gamma$, and  its branching ratio is given by
 \begin{align}
 BR(\ell_j\to \ell_\alpha\gamma)\approx \frac{48\pi^3\alpha_{em} C_{j} }{G_F^2 m_{\ell_j}^2}|a_{j\alpha}|^2
 \end{align}
 where $\alpha_{em}\approx1/137$ is the fine-structure constant,
$C_j=(1,1/5)$ for ($i=\mu,\tau$), ${G_F}\approx1.17\times 10^{-5}$ GeV$^{-2}$ is the Fermi constant, $a_{j\alpha}$ is
computed as
\begin{align}
& a_{j\alpha}=
-\frac{s_R c_R }{2\sqrt2(4\pi)^2}
\sum_{k=1}^2
(y'_\ell)_{\alpha k} m_{L'_k} (y_{L'})_{k,j} 
\left[F_{\ell_j\to\ell_\alpha\gamma}(m_{H_1},m_{L'_k}) - F_{\ell_j\to\ell_\alpha\gamma}(m_{H_2},m_{L'_k}) \right],\\
%%%
& F_{\ell_j\to\ell_\alpha\gamma}(m_{1},m_{2})
 =
 \frac{2 m_1^4-4 m_1^2 m_2^2 + m_2^4+4 m_1^4 \ln\left[\frac{m_2}{m_1}\right]}{2(m_1^2-m_2^2)^3}.
\end{align} 

%%%%%%%%%%%%%%%%%
{\it Muon anomalous magnetic moment $(g-2)_{\mu}$}:
Through the same process from the above LVFs, there exists the contribution to $(g-2)_{\mu}$, and 
its form $\Delta a_\mu$ is simply given by
\begin{align}
\Delta a_\mu \approx -\frac{m_\mu a_{22}}{2}.
\end{align}
This value can be tested by current experiments~\cite{bennett, discrepancy1, discrepancy2}.

%%%%%%%%%%%%%%%%%
{\it $\mu-e$ conversion}:
%%%
The $\mu-e$ conversion process can be found in the same diagram as the process of $\ell_j\to \ell_\alpha\gamma$ with $\gamma$ line being attached to nucleons, where additional contribution is taken into account by replacing $\gamma$ with $Z$ boson.
Then the $\mu-e$ conversion rate $R$
 is given by~\cite{Hisano:1995cp}
\begin{align}
&R=\frac{\Gamma(\mu\to e)}{\Gamma_{\rm capt}}\\
&=
%4\alpha_{\rm em}^5 \frac{Z^4_{\rm eff}|F(q)|^2 m^5_\mu}{Z}
\frac{C_{\mu e}}{\Gamma_{\rm capt}}
\left|Z\left(b_{21}^\gamma-\frac{a_{21}}{m_{\mu}}\right)
- b_{21}^Z \frac{(2Z+N)A_u+(Z+2N)A_d}{2(s_{tw}c_{tw})^2} \right|^2,\\
%%%
&b^V_{21}=
\frac{s_R c_R }{2\sqrt2(4\pi)^2}
\sum_{k=1}^2
(y'_\ell)_{\alpha k} m_{L'_k} (y_{L'})_{k,j} 
\left[F_{\mu e}(m_{H_1},m_{L'_k},m_V) - F_{\mu e}(m_{H_2},m_{L'_k},m_V) \right],\\
%%%
&F_{\mu e}(m_1,m_2,m_3)=
\int_0^1dx_3\int_0^{1-x_3}dx_4 \frac{x_4(1-x_4)}{x_3 m_1^2 +(1-x_3) m_2^2+x_4(x_3 + x_4 - 1) m_3^2},
%\int_0^1dx \int_0^{1-x}dz \frac{z(1-z)}{x m_1^2 +(1-x) m_2^2+z(x + z - 1) m_3^2},
\end{align}
where $V\equiv (\gamma, Z)$, and $m_\gamma=0$, and $m_Z\approx 91.19$ GeV, $C_{\mu e}\equiv4\alpha_{\rm em}^5 \frac{Z^4_{\rm eff}|F(q)|^2 m^5_\mu}{Z}$, $A_u\equiv -\frac12-\frac43s_{tw}^2$, $A_d\equiv -\frac12+\frac23s_{tw}^2$, $\sin^2\theta_w\equiv s_{tw}^2\approx0.23$. The values of $\Gamma_{\rm capt}$, $Z$, $N$, $Z_{\rm eff}$, and $F(q)$ depend on the kind of nuclei. Here  we focus on Titanium, because its sensitivity will be improved by several orders of magnitude~\cite{Hungerford:2009zz, Cui:2009zz} in near future compared to the current bound~\cite{Dohmen:1993mp},
as can be seen in the Table~\ref{tab:Cif}.
In this case,  these values are determined by $\Gamma_{\rm capt}=2.59\times10^6$ sec$^{-1}$, $Z=22$, $N=26$, $Z_{\rm eff}=17.6$,  and $|F(-m_\mu^2)|=0.54$~\cite{Alonso:2012ji}.~\footnote{Notice here that all the contributions discussed in Figs.~1 in this paper are negligible in our model. Especially the box diagrams, which consists of two contributions, cancels each other, when running  fermions in the loop are active neutrinos only, which are almost massless compared with W boson. }

\section{Numerical analyses}
%First of all we fix $\lambda_{\Phi S}=0.05$ for simplicity in the numerical analysis.
Now that we have all the formulae for the quark and lepton sector, and we carry out numerical analysis to find what kind of regions
are allowed. Here we randomly select values of the fifteen parameters within the corresponding ranges
\begin{align}
& m_{\eta_I} \in [250\ \text{GeV}, 500\ \text{GeV}],\quad
 m_{H_1} \in [600\ \text{GeV}, 800\ \text{GeV}],\quad
 m_{H_2} \in [4\ \text{TeV}, 6\ \text{TeV}],\nn\\
& m_{Q'_1} \in [4.5\ \text{TeV}, 5\ \text{TeV}],\quad m_{Q'_2} \in [1.7\ \text{TeV}, 2.2\ \text{TeV}],\\
& m_{L'_1} \in [7\ \text{TeV}, 7.5\ \text{TeV}],\quad m_{L'_2} \in [9\ \text{TeV}, 10\ \text{TeV}],\\
%%%
&\{ (y_u)_{12}, (y_u)_{22} \} \in [-1,1],\quad 
(y_u)_{11} \in [-0.1,0.1], \quad (y_u)_{21} \in [-0.002,0.002],
%%%
\nn \\
%%%
&\{ (y_\nu)_{21}, (y_\nu)_{22} \} \in [-7\times10^{-12},7\times10^{-12}],\quad 
(y_\nu)_{11} \in [-3\times10^{-13},3\times10^{-13}], \nn \\
& (y_\nu)_{12} \in [-2\times10^{-13},2\times10^{-12}],
	\label{range_scanning}
\end{align}
%%%
to reproduce quark masses, CKM mixings for the quark sector, and neutrino oscillation data and 
satisfy the constraints of LFVs for the lepton sector. In this analysis, we are preparing 10 million sample points. 
Notice here that
%, applying the Yukawa couplings in Eq.~(\ref{range_scanning}) as inputs,
the other Yukawa couplings such as $y_Q$, $y_d$, $y_L$, $y_\ell$ are numerically solved by using the best fit values of the measurements in ref.~\cite{pdg} for quark sector and ref.~\cite{Forero:2014bxa} for lepton sector.
Then we obtain the sets of Yukawa couplings where 
 all we need to take care is not to exceed the perturbative limit that we take $\sqrt{4\pi}$ as upper limit. 
The sets of Yukawa couplings are applied to calculate $\Delta m_D$, $\Delta m_K$,  $\Delta a_\mu$, $\ell \to \ell' \gamma$ and $\mu-e$ conversion.

\begin{figure}[t]
\begin{center}
\includegraphics[scale=0.6]{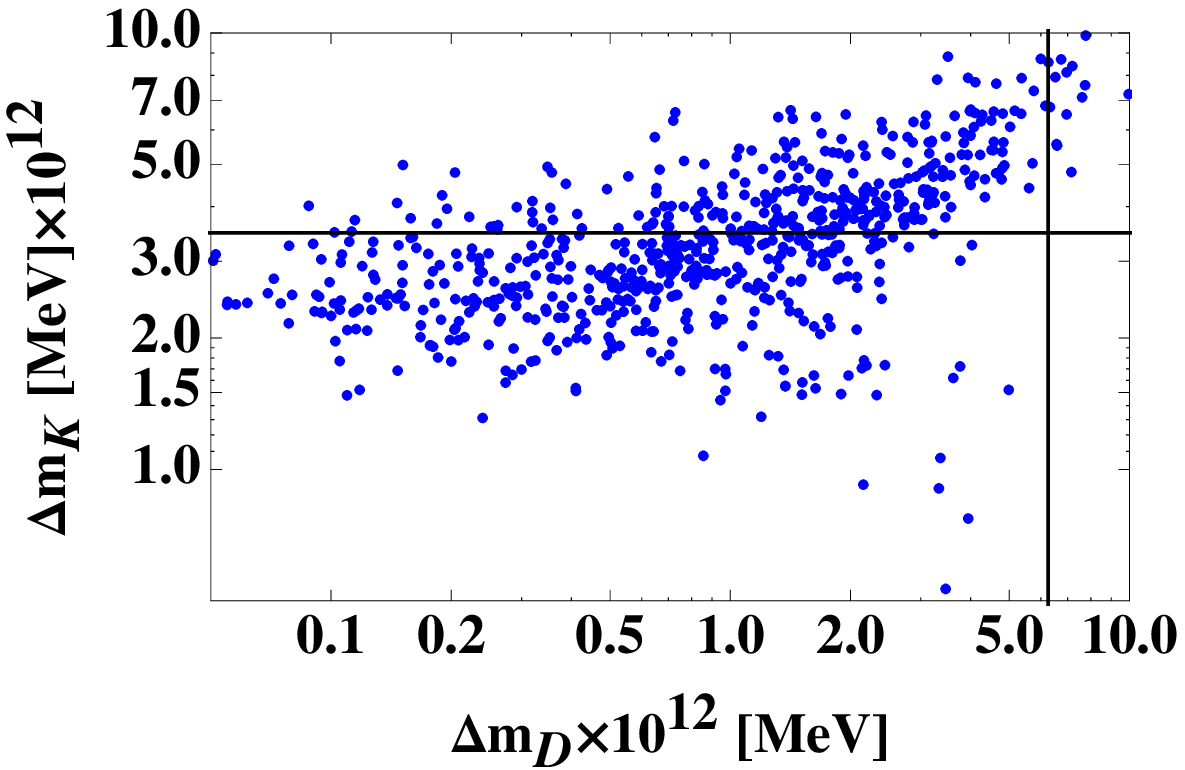}
%%%
\includegraphics[scale=0.6]{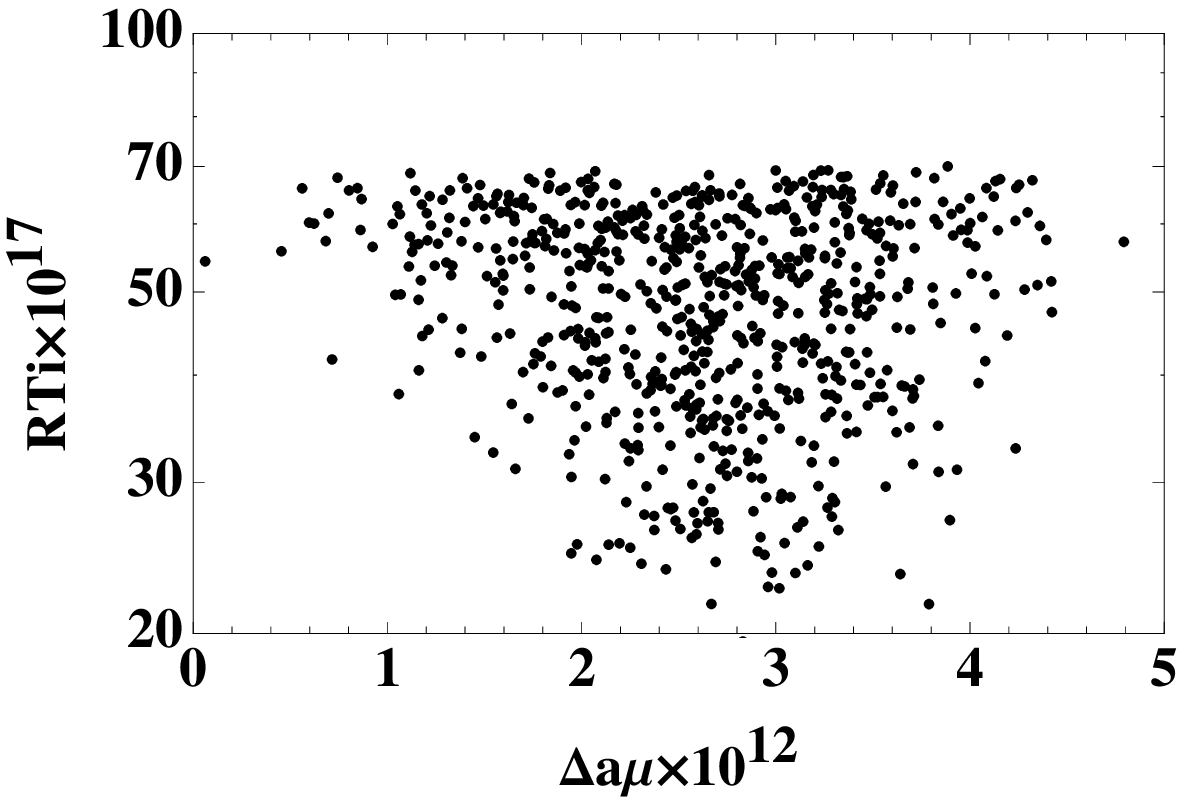}
\caption{The left figure is the scattering plot in terms of $\Delta m_D$ and $\Delta m_K$ normalized by [MeV]$\times 10^{12}$, and the black solid lines represents the experimental upper bound. The right figure is the scattering plot in terms of $\Delta a_\mu\times 10^{12}$ and $R_{Ti}\times 10^{17}$.
}
\label{fig:results}
\end{center}
\end{figure}

{\it Numerical results}:
The numerical results are shown in Fig.~\ref{fig:results}. 
The left figure is the scattering plot in terms of $\Delta m_D$ and $\Delta m_K$ normalized by [MeV]$\times 10^{12}$, and the black solid lines represents the experimental upper bound.
Here we have found 734 allowed points, and the most stringent constraint comes from the process of $\mu\to e\gamma$.
%It tells us that $\Delta m_K$ is completely safe because the maximal allowed region is always lower than the experimental bound by one order of magnitude. 
{ It tells us that our $\Delta m_K$ values are the same order as current experimental constraint and some parameter sets are excluded.
Similarly $\Delta m_D$ values can be comparable to the current experimental bound. Thus they can be tested in the near future. }
%%%
 The right figure is the scattering plot in terms of $\Delta a_\mu\times 10^{12}$ and $R_{Ti}\times 10^{17}$.
It tells us that the maximal value for $(g-2)_\mu$ is around $5\times 10^{-12}$, which is lower than the current bound by three order of magnitude. $R_{Ti}$ is also much lower than the current bound, however it will be test in the future experiment such as COMET~\cite{Hungerford:2009zz, Cui:2009zz}, which will reach $R\approx10^{-18}$  as shown in the previous section. 
%%%

%%%%%%%%%%%%%%%%%%%
%%%%%%%%%%%%%%%%%%%

%\section{Collider physics}
%\textcolor{red}{If you like, please add.}

\section{ Conclusions and discussions}
We have proposed a radiatively induced quark and lepton mass model in the first and second generation, in which we have analyzed the allowed regions simultaneously to satisfy the FCNCs for the quark sector and LFVs including $\mu-e$ conversion in addition to the quark mass and mixing and the lepton mass and mixing. Also we have estimated the typical value for the $(g-2)_\mu$.
%%%

{Then we have found $\Delta m_K$ and $\Delta m_D$ can be the same order as the current experimental bound where some parameter sets are excluded.
Thus our model can be further tested in the near future.}
%is completely safe because the maximal allowed region is always lower than the experimental bound by one order of magnitude. While $\Delta m_D$ can be comparable to the current experimental bound. Thus it can be tested in the near future. 
%%%
As for the lepton sector, we have found that the maximal value for $(g-2)_\mu$ is around $5\times 10^{-12}$, which is lower than the current bound by three order of magnitude. $R_{Ti}$ is also much lower than the current bound, however it will be test in the future experiment such as COMET, which will reach $R\approx10^{-18}$. 

{We note that the $Z'$ boson from $U(1)_R$ gauge symmetry has flavor violating interaction due to our choice of charge assignment for SM fermions. 
Since $Z'$ couples to both SM quarks and leptons these interaction could be tested in future LHC experiments. 
Particularly lepton flavor violating signals $pp \to Z' \to \ell \ell'$ would be interesting signatures of the model.
Detailed simulation studies of the signal is beyond the scope of this paper and we left it as a future work.
}

At the end of this paper, we mention the dark matter candidate.
In our case (and our parametrization), ${\eta_I}$ can be a dark matte candidate, which is the imaginary component of the $SU(2)_L$ doublet inert boson.
The dominant annihilation processes are induced through the gauge interactions, since Yukawa couplings related to $\eta$, $y_\nu$ and $y_\ell$, are expected to be tiny.  Thus its nature is the same as the two Higgs doublet model with one inert boson and serious analysis can be found in ref.~\cite{Hambye:2009pw, LopezHonorez:2006gr}.
% Thus we do not discuss the dark matter sector in this paper anymore.
It can also be detected through the spin independent direct detection searches such as LUX~\cite{Akerib:2013tjd}, because it has two Higgs portal interactions with the nucleon. This  situation might relax the experimental constraint compared to the one Higgs portal scenario, activating the cancellation mechanism between two CP-even bosons~\cite{Baek:2012uj}.

%\section*{ Appendix}
%%%%%%%%%%%%%%%%%%%...

%\newpage
%%%%%%%%%%%%%%%%%%%%%%%%%%%%%%%%%%%
%\hspace{0.2cm} {\bf Acknowledgments}
%\section*{Acknowledgments}:
%\vspace{0.5cm}
\section*{Acknowledgments}
\vspace{0.5cm}
H. O. is sincerely grateful for all the KIAS members, Korean cordial persons, foods, culture, weather, and all the other things.
%%%%%%%%%%%%%%%%%%%%%%%%%%%%%%%%%%%
%%%%%%%%%%%%%%%%%%%%%%%%%%%%%%%%%%%

\end{document}